\documentclass[8.5pt,twoside,twocolumn]{article}
\usepackage[super,sort&compress,comma]{natbib}
\usepackage{mhchem}
\usepackage{sectsty}
\usepackage{balance}
\usepackage{amssymb}

\usepackage{graphicx} 
\usepackage{lastpage}
\usepackage[format=plain,justification=raggedright,singlelinecheck=false,font=small,labelfont=bf,labelsep=space]{caption}
\usepackage{fancyhdr}
\begin{document}

\thispagestyle{plain}
\fancypagestyle{plain}{
\fancyhead[L]{\includegraphics[height=8pt]{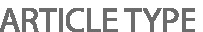}}
\fancyhead[C]{\hspace{-1cm}\includegraphics[height=20pt]{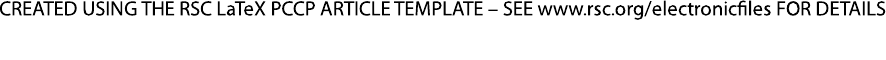}}
\fancyhead[R]{\includegraphics[height=10pt]{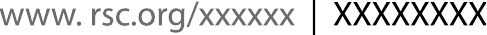}\vspace{-0.2cm}}
\renewcommand{\headrulewidth}{1pt}}
\renewcommand{\thefootnote}{\fnsymbol{footnote}}
\renewcommand\footnoterule{\vspace*{1pt}%
\hrule width 3.4in height 0.4pt \vspace*{5pt}}
\setcounter{secnumdepth}{5}

\makeatletter
\def\subsubsection{\@startsection{subsubsection}{3}{10pt}{-1.25ex plus -1ex minus -.1ex}{0ex plus 0ex}{\normalsize\bf}}
\def\paragraph{\@startsection{paragraph}{4}{10pt}{-1.25ex plus -1ex minus -.1ex}{0ex plus 0ex}{\normalsize\textit}}
\renewcommand\@biblabel[1]{#1}
\renewcommand\@makefntext[1]%
{\noindent\makebox[0pt][r]{\@thefnmark\,}#1}
\makeatother
\renewcommand{\figurename}{\small{Fig.}~}
\sectionfont{\large}
\subsectionfont{\normalsize}

\fancyfoot{}
\fancyfoot[LO,RE]{\vspace{-7pt}\includegraphics[height=9pt]{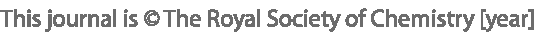}}
\fancyfoot[CO]{\vspace{-7.2pt}\hspace{12.2cm}\includegraphics{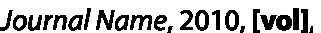}}
\fancyfoot[CE]{\vspace{-7.5pt}\hspace{-13.5cm}\includegraphics{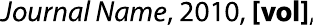}}
\fancyfoot[RO]{\footnotesize{\sffamily{1--\pageref{LastPage} ~\textbar  \hspace{2pt}\thepage}}}
\fancyfoot[LE]{\footnotesize{\sffamily{\thepage~\textbar\hspace{3.45cm} 1--\pageref{LastPage}}}}
\fancyhead{}
\renewcommand{\headrulewidth}{1pt}
\renewcommand{\footrulewidth}{1pt}
\setlength{\arrayrulewidth}{1pt}
\setlength{\columnsep}{6.5mm}
\setlength\bibsep{1pt}

\twocolumn[
  \begin{@twocolumnfalse}
\noindent\LARGE{\textbf{Static friction scaling of physisorbed islands: the key is in the edge}}
\vspace{0.6cm}

\noindent\large{\textbf{Nicola Varini,\textit{$^{a,b}$} Andrea Vanossi,\textit{$^{c,d}$}
Roberto Guerra,\textit{$^{d,c}$}
Davide Mandelli,\textit{$^{d}$}
Rosario Capozza,\textit{$^{d}$}
and Erio Tosatti\textit{$^{\ast,d,c,e}$}}}

\vspace{0.5cm}

\noindent\textit{\small{\textbf{Received Xth XXXXXXXXXX 20XX, Accepted Xth XXXXXXXXX 20XX\newline
First published on the web Xth XXXXXXXXXX 200X}}}

\noindent \textbf{\small{DOI: 10.1039/b000000x}}
\vspace{0.6cm}

\begin{abstract} 
The static friction preventing the free sliding of nanosized rare gas solid islands physisorbed on incommensurate crystalline surfaces
is not completely understood. Simulations modeled on Kr/Pb(111) highlights the importance and the scaling behavior of the island's edge 
contribution to static friction.

\end{abstract}

\vspace{0.5cm}
 \end{@twocolumnfalse}
  ]

\footnotetext{\textit{$\ast$}~tosatti@sissa.it}
\footnotetext{\textit{$^{a}$~Research \& Development, Curtin University, GPO Box U 1987, Perth 6845, Australia.}}
\footnotetext{\textit{$^{b}$~iVEC, 26 Dick Perry Avenue, Kensington WA 6151, Australia.}}
\footnotetext{\textit{$^{c}$~ CNR-IOM Democritos National Simulation Center,Via Bonomea 265, I-34136 Trieste, Italy.}}
\footnotetext{\textit{$^{d}$~International School for Advanced Studies (SISSA), Via Bonomea 265, I-34136 Trieste, Italy.}}
\footnotetext{\textit{$^{e}$~International Centre for Theoretical Physics (ICTP), Strada Costiera 11 I-34151 Trieste, Italy.}}

\section{\bf{Introduction}}

\noindent \normalsize{Static friction -- sometimes desirable, often a nuisance, always important -- is less
studied than its more famous counterpart, dynamic friction. A standing puzzle
is the persistence of static friction even for ideal mesoscale sized sliders such as incommensurate
physisorbed islands on atomically perfect crystal surfaces, where one could expect it to asymptotically vanish. Here we
show, based on prototype atomistic simulations of rare gas islands sliding on a crystal surface,
that the slider's {\it edges} represent the ultimate actors. When the island body is ready to slide
``superlubrically'', the sliding is impeded by an edge-originated barrier that prevents the misfit dislocations or solitons --
tiny density and corrugation modulations with the beat periodicity between adsorbate and surface --
from moving. Only when the static friction force is reached, the barrier vanishes locally at one point on the edge,
solitons enter through this point to sweep the island, which only then becomes depinned and free-sliding.
We show that finite size scaling of static friction of a superlubric rare gas island with area $A$ has the same form  $A^{\gamma_s}$
as discussed in literature for dynamic friction, but with specific, generally distinct static scaling exponents $\gamma_s$ that are
edge-controlled and smaller than 1/2, as opposed to $\gamma_s=1$ for pinned, commensurate islands.
The island static friction connection with edges, here addressed realistically in rare gas islands for the
first time, constitutes the controlling factor for inertial depinning in quartz crystal microbalance
experiments, and is expected to be a factor of importance for nanotribological applications
involving weak contacts.}

Understanding the atomic-scale mechanisms which control friction is probably one
of the most important challenges in nanotechnology\cite{bhushan2002,urbakh2004,urbakhmeyer,
vanossi2013}. Static friction -- the threshold force necessary to initiate the sliding between two contacting
surfaces, which Coulomb distinguished historically from dynamic friction -- is basic not just to our everyday
experience, including standing and walking, but is
crucial, and often fatal, to the working of nanosystems\cite{frenken2006}. In nanomotive devices,
as well as in elementary processes such as the pulling or pushing of nano-objects on a surface by a tip\cite{schirmeisen08,dietzel13,
kawai2014}, or by a colliding molecule\cite{allison2009}, or by simple inertia\cite{krim_review,krim_advphys},
static friction constitutes the ultimate obstacle that affects the onset of  sliding.
Often undesired, static friction of nano-sliders can also  be of help, e.g., to prevent the diffusive sliding
away of adsorbed molecules, small proteins clusters and of islands, favoring so-called nano-positioning
when deposited on surfaces\cite{junno,sitti}.
Despite its importance, and several notable exceptions\cite{ hurtadoI99, hurtadoII99, robbins1999, muser01, muser03, gao10},
static friction is not as much discussed and qualified as dynamic friction,
which is generally much more popular.
Differently from macroscopic sliding where relatively complex processes
due to asperities and multiple contacts \cite{persson_book}, or to load inhomogeneity\cite{ hurtadoI99, hurtadoII99, gao10} 
are at play, homogeneous crystalline nano-sliders offer a unique
opportunity to address static friction in its simplest form -- that between perfectly regular, periodic faces.

Given two crystals in contact, e.g., a two-dimensional (2D) adsorbate monolayer island (the slider)
and a perfect crystal surface (the substrate), different possibilities may occur. The island lattice may
be commensurate (C) with the surface lattice, or it may be incommensurate (IC), meaning that the two
lattice 2D cells cannot be made to coincide by any rational scale change $p/q$, where $p$ and $q$ are two integers. 
In the general C case, and also in the IC case if the slider is ``soft'' (inter-atomic forces weaker than or similar to the typical substrate
corrugation forces, causing effective commensurability to prevail), the island will exhibit static friction, due to pinning of the two lattices.
The latter situation of strong contact has been theoretically addressed by a number of workers\cite{hurtadoI99, hurtadoII99, robbins1999, muser01, muser03, gao10, zapperi, dellago13}.
Conversely, static friction should ideally vanish, foreshadowing free sliding, also called ``superlubricity'',
for a prototype weak contact, such as an infinite, hard IC crystalline slider, weakly interacting with a perfectly periodic substrate\cite{peyrard83,hirano90,shinjo93,
cieplak94,braun_book, note1}.

Experimentally, very low friction dynamic
sliding has been observed, e.g., in graphitic materials by extracting
telescopic carbon nanotubes\cite{superlubric_china, superlubric_bocquet} or by rotating
out of registry an AFM-tip driven graphite flake on a graphite substrate\cite{dienwiebel04,dienwiebel08,fasolino2013}.
Evidence of low dynamic friction in noncrystalline sliding has been collected by
AFM manipulation of amorphous and crystalline antimony nanoparticles deposited on a graphite
substrate\cite{schirmeisen08,schirmeisen09,dietzel13}. At the same time however the
corresponding static friction, even if conceptually and practically different, and no less important,
was generally skirted in these low-friction contacts.
In recent studies of 2D incommensurate colloidal crystals
sliding over optical lattices, static friction emerged clearly, along with a pinned-to-superlubric
transition for decreasing periodic potential magnitudes\cite{Bohlein12,Vanossi12PNAS,Vanossi12}.
Static friction controls directly the quartz crystal microbalance (QCM) sliding experiments
\cite{krim_review,krim_advphys, mistura0} which we consider as our reference system. In QCM a crystal surface
with a layer, but particularly a submonolayer coverage of physisorbed inert gas is periodically shaken to dislodge the
2D crystalline islands formed by coalescence of the adsorbed gas molecules, giving rise to inertial
sliding friction. These submonolayer islands generally constitute incommensurate sliders\cite{zeppenfeld92},
which as a rule should behave as hard (even if not at all rigid) since the gas interatomic energy, of order 15 meV, although
much weaker than the atom-surface adhesive energy of order 150 meV,  is usually much stronger than
the substrate corrugation, typically only a few meV\cite{Bruch07}. QCM data in UHV generally show that for low
rare gas coverage, low temperature, and moderate oscillatory force magnitude, adsorbed islands 
remain 
stuck to the
surface and do not slide, despite cleanliness, hardness and incommensurability\cite{mistura1,mistura0, mistura2, bruschi10}.
The islands remain pinned by static friction until either the growing inertial force, or the increasing
coverage -- which implies a larger average island size\cite{Park99} 
and thus a larger inertial force -- 
or temperature, reach a threshold value.  Only beyond that point,
the static friction is overcome, the islands  depin, and inertial sliding takes place, with a nonzero slip time whose
inverse measures dynamic friction\cite{krim_review,krim_advphys, mistura0}.

Our main goal is to understand the ultimate reason why static friction 
should
persist even for ideally perfect
rare gas islands adsorbed on crystalline surfaces, which constitute weak contacts where owing to incommensurability
free superlubric sliding could be expected. Real-life surfaces are of course far from perfect, so that an adsorbed island
will generally attach to surface steps, defects, impurities, etc., which can provide extrinsic pinning\cite{robbins1999}.
Nevertheless, static friction of cold adsorbed islands generally survives the progressive elimination
of defects, 
and their pinning effect by reducing the oscillation amplitude\cite{renner01,fois07, bruschi10}. For a defect free, genuinely incommensurate hard
island, such as is realized for example by Kr/Pt(111)\cite{kern87}, and most other
cases where the island-surface commensurability ratio is known to drift continuously with temperature,
there should be no barrier and no pinning. Here an intrinsic source of static friction must be at work.

The edge is an intrinsic ``defect'' that every island, cluster, or deposited nanosystem,
must have. We conducted simulations mimicking realistic 2D incommensurate rare
gas islands up to very large size, adsorbed on defect free metal surfaces. These simulations clearly show that
even without defects, these weak contacts exhibit a basic static friction threshold dictated by the island edges.
Specifically, for an adsorbed island, where the misfit lattice-mismatched dislocations (i.e.\ solitons) pre-exist, 
we find that the entry through the island edge of a new soliton is the event that
initiates the depinning and the subsequent superlubric sliding. Soliton entry however is not cost free;
the pushing force must overcome an edge-related energy barrier, which is thus the controlling element
of the island's static friction. The barrier's relative role and importance, and the ensuing static friction
are found to decrease with increasing island size and temperature, precisely as seen in experiments. Since
the static friction is edge-originated, its scaling $F_s  \sim A^{\gamma_s}$ with the island's area $A$
is not only sub-linear, $\gamma_s < 1$, but sub-linear even with respect to the coarse-grained island perimeter $2 (A\pi)^{1/2}$, 
that is $\gamma_s < 1/2$,
indicating that only a zero-measure subset of edge points is responsible for the pinning barrier.  
By contrast with this result for incommensurate islands we examine in parallel that of commensurate
islands, strong contacts which have a ``bulk'' pinning, with a trivial area scaling exponent $\gamma_s$\,=\,1.
Even there nonetheless the edge plays a role. We find that the lattice dislocation triggering the sliding
of a commensurate island nucleates, as  also discussed by previous workers 
in different situations\cite{hurtadoI99, hurtadoII99, gao10},
preferentially at the island edge, here acting therefore as a facilitating element which quantitatively reduces the
static friction.

In the far more common case (for physisorption) of incommensurate islands, we will identify, at the general level, the emerging
key parameters which control edge-induced static friction. They are: a) the shape, rugosity, and elasticity of the island edge, which
influence $\gamma_s$, a sublinear size scaling exponent that does not appear to exhibit universality; b) the amplitude of
the substrate corrugation potential, which controls the static friction magnitude; c) the temperature, influencing
static friction both directly, and indirectly through the island's changing incommensurability.

The role of edge-induced static friction is, we propose, especially important for QCM experiments,
a context in which it has apparently not been previously discussed. As was said above, it is a general
observation in QCM that lack of inertial sliding, and a consequent zero slip time, occurs in all
cases for a sufficiently low coverage and low temperature, a regime where the adsorbate forms
crystalline islands that are generally incommensurate with a crystalline substrate\cite{zeppenfeld92, renner01,  mistura0,
fois07, bruschi10, krim_review,krim_advphys}.
We will show that realistic parameters for the adsorbed rare gas on a metal surface predict that the
QCM inertial force  cannot reach, as coverage increases, the low temperature static friction threshold for
depinning until the island diameters grow as large as 
many tens when not a hundred of nanometers. 
That is a remarkably large size, comparable to that of terraces or facets of even a good quality metal surface. The conclusion is
thus that intrinsic edge pinning is not just qualitatively, but quantitatively important in QCM and other
incommensurate nano-sliding experiments. We also find that the effects
of temperature are diverse and intriguing. Besides increasing the edge rugosity, temperature does
reduce the effective substrate corrugation, which in turn lowers the static friction threshold.
At the same time, the commensurability ratio of the two lattices may drift with temperature because the island's
thermal expansion is generally unmatched by that of a metal substrate. As temperature grows, one
or more low order commensurabilities may accidentally be hit, at which point a geometrical interlocking
barrier against sliding arises, leading in principle to the novel possibility of reentrant static friction.

\section {\bf MD simulations of rare gas islands and their static friction on a metal surface: Kr/Pb(111)}

To substantiate our discussion with a specific case study, we model an incommensurate rare gas island on a metal surface
with parameters appropriate to Kr/Pb(111), a prototype system of current interest\cite{fois07} which we adopt here
as a generic model hard slider. The atoms interact mutually via Lennard-Jones forces
(standard parameters given in Methods). We restrict to temperatures well below half the adsorbate bulk
melting temperature, $T \ll T_m/2$ (for $T_m = 115$\,K, we employ $T \ll 60$\,K), where the adsorbate island
does not yet melt, remaining crystalline with a 2D triangular lattice only weakly modulated by the incommensurate
and essentially rigid underlying (111) metal substrate. For different assumed shapes we simulate a sequence of
islands consisting of $N$ adsorbate atoms, with $N$ increasing up to a maximum size of about $300000$.
Using molecular dynamics we first relax and anneal the islands -- initially cut out of a perfect triangular lattice  -- in
the (111) periodic potential mimicking the perfect crystal substrate, and optimize the overall island adsorption geometry
so as to minimize energy. In the resulting zero temperature state, the 
main interior part of the island 
and the substrate are incommensurate,
with the model Kr atom-atom spacing $a=4.057$\,\AA~only slightly smaller than $(7/6) a_s = 4.084$\,\AA,
where $a_s = 3.50$ \,\AA~is the nearest neighbor distance in Pb(111).

\begin{figure*}
\centering
\includegraphics[angle=0,width=0.8\textwidth]{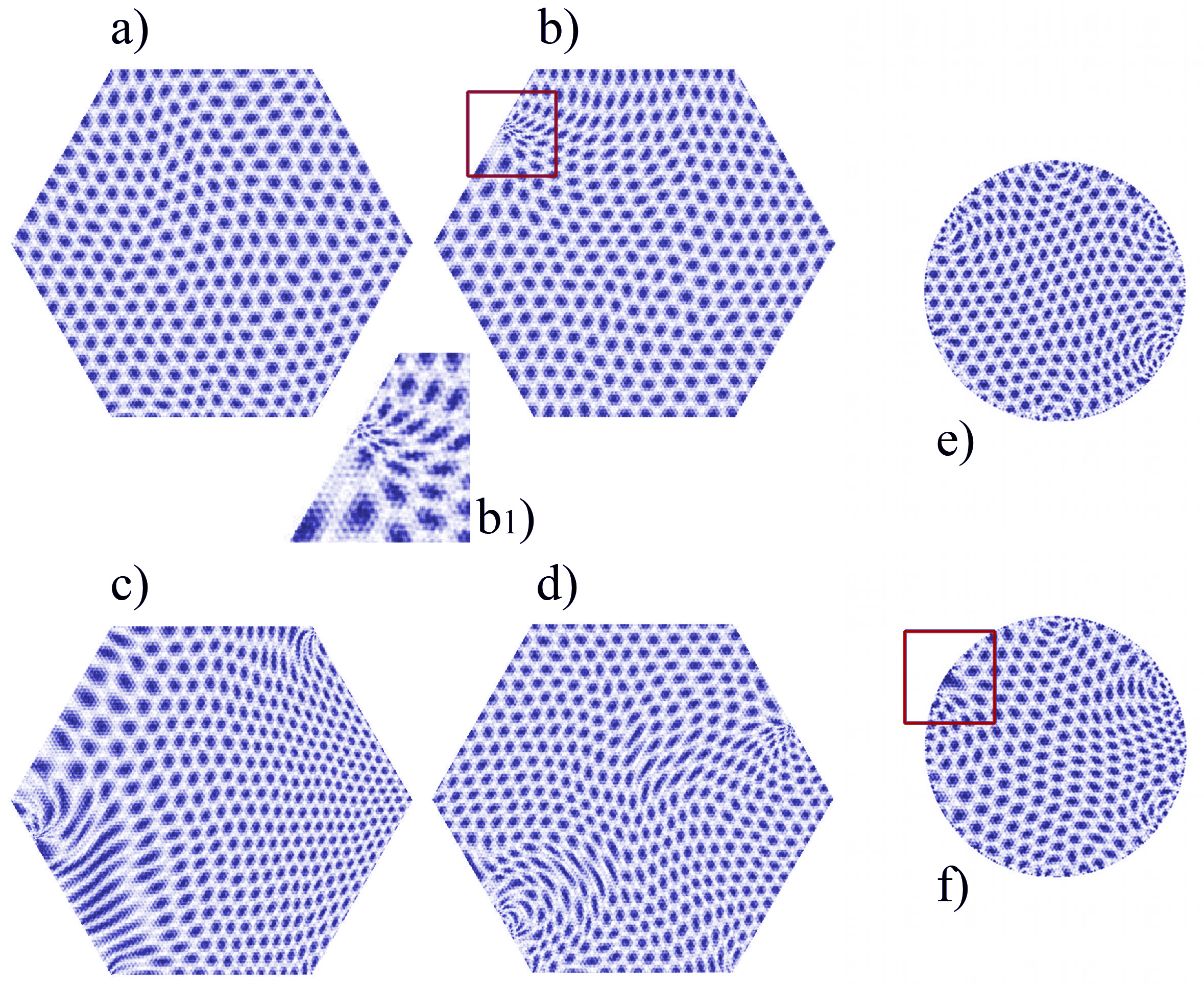}
\caption{\label{solitoni} {Model ideal hexagonal ($N=269101$) and circular ($N=177507$)
Kr islands adsorbed on Pb(111). Pictures show maps of the island, with colors ranging from dark blue
when Kr atoms are maximally coincident with a (6/7) rescaling of the underlying Pb atom positions, to clear
when they are minimally coincident. The resulting Moir\'e patterns, enhanced in this manner for visibility (see Methods)
highlight the soliton network between the island and surface lattices and their evolution
at the static friction threshold. (a) After annealing, at zero temperature and zero applied force ($T=0$,
$F_{ext}=0$); (b) with applied force $F_{ext}=1.4$\,eV/\AA, just above the static friction value,
and right after the soliton entry at a left edge corner -- highlighted by the red square and magnified
by the zoomed in region (b1); (c) same system after sliding of the island center by one surface
lattice spacing.
Note the density accumulation at the front edge and rarefaction at the trailing edge, showing the frictional role of the edges;
(d) same system at later time when the sliding is 1.5 lattice spacings, and a soliton
exits the island on the right hand side. (e) Circular island, just after annealing ($T=0$, $F_{ext}=0$); 
(f) with applied $F_{ext}$ slightly above the static friction threshold. Here
again the static friction is determined by the entry of a soliton at the left edge, highlighted by the red square.}}
\end{figure*}

The surface-deposited islands develop delicately regular soliton superstructures, 
representing the deviation from exact (6/7) commensurability, as 
pictured in  blue scale
in Fig.~\ref{solitoni} by means of a contrast enhancement technique (see Methods).
These pictures highlight the great deformability of this gossamer superstructure, well beyond any rigid or nearly
rigid approximation. Compressional/dilational strains in the 2D lattice are directly reflected by increased/decreased
density of superstructures.

\section{\bf {Island depinning and  size scaling of static friction}}

We subsequently simulate the forced depinning of the island (previously prepared at a desired temperature
using a thermostat, then switched off,  see Methods), by applying to the island center of mass
a constant planar total force $F_{ext}$,
that is by applying to each Kr atom a force $F_{ext}/N$. 
In our protocol the static friction is the lowest value of $F_{ext}$
sufficient to cause, within 1.2 ns simulation time, a center-of-mass drift of two substrate
interatomic spacings, or 0.7 nm,  signaling the depinning of the island\cite{QCMnote}. The shapes of the
island edge, in principle quite important for static friction, are not easily equilibrated in MD simulations at the
low temperatures considered. We therefore examine two ideally opposite test shapes, perfect hexagonal and
circular -- neither of them realistic but providing together a fair idea of the generic behaviour to be expected.
Torques and sliding-induced overall rotations arising from possible non-centered island shapes are also neglected.
Despite the natural random asymmetry expected of realistic islands, their large size prevents 
the brownian motion of their orientations (as well as of their centers of mass) making the torque needed 
to cause their overall rotation generally very large for QCM.  
%

\begin{figure}[t!]
\centering
\includegraphics[angle=0,width=8.5cm]{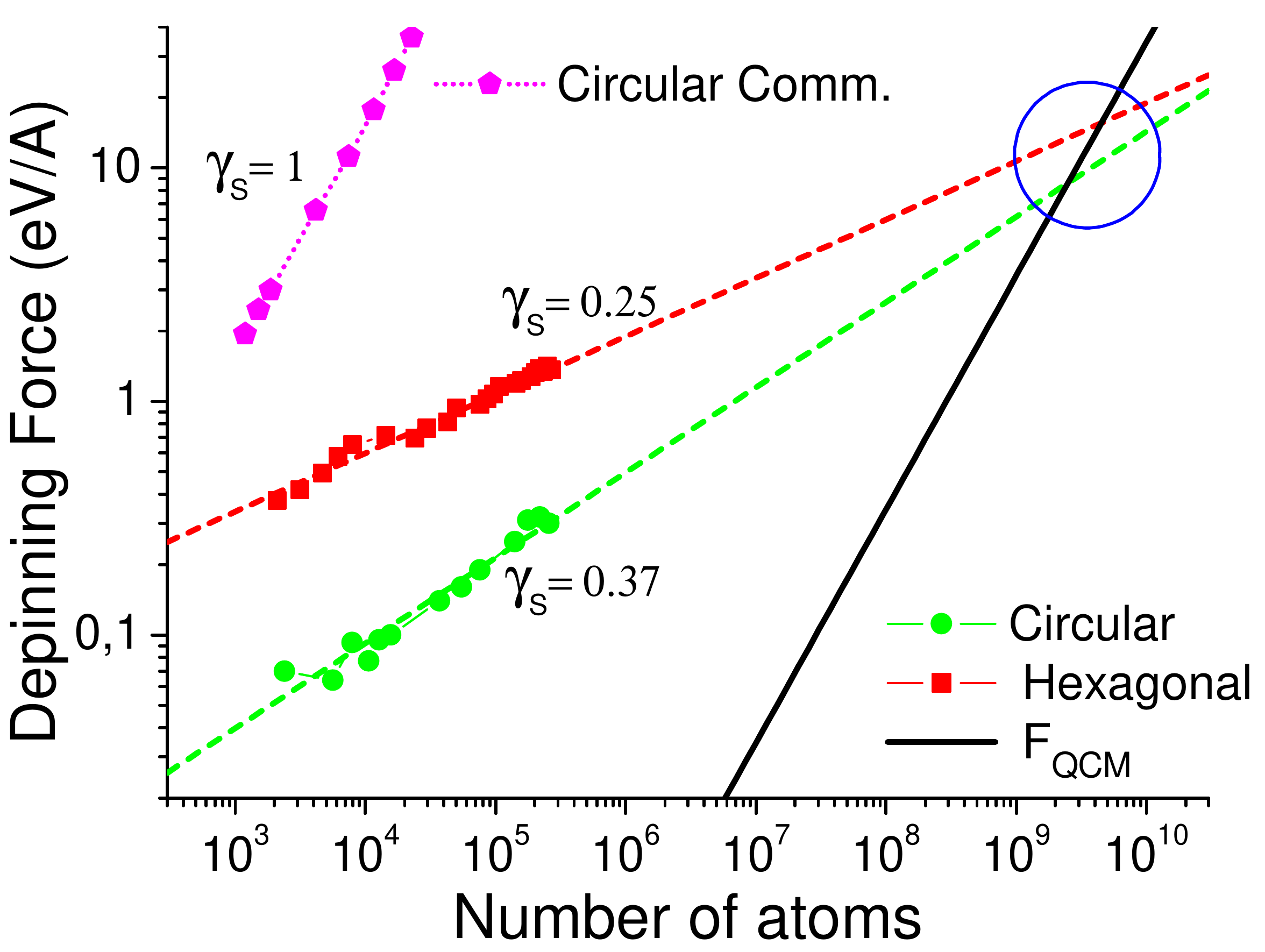}
\caption{\label{depforce} { Size scaling of total static friction force $F_s$ of simulated incommensurate
adsorbed islands (modeled on Kr/Pb(111)) at $T$\,=\,0\,K
as a function of the atom number $N$ (log-log scale) for {\it i}) hexagonal islands (squares); {\it ii)} circular
islands (circles). Assumed surface corrugation: 5\% of the atom adsorption energy; note the strongly sublinear scaling for both shapes.
The intersection with the estimated QCM inertial force (black solid line) is emphasized (blue circle), showing
how the edge-originated  static friction alone can prevent island sliding up to
large island sizes, which qualitatively corresponds to large sub-monolayer adsorbate coverages.
By comparison we also show the size scaling of static friction ($\gamma_s$\,=\,1) for
a {\it commensurate} adsorbate at $T$\,=\,50\,K (see Methods).
The commensurate frictional stress $F_s/A$ is about $10^7 N/m^2 $  compared with
a much smaller 40 $N/m^2 $ of incommensurate islands at inertial depinning.}}
\end{figure}

Fig.~\ref{depforce} shows, in log-log scale, the overall static friction force $F_0(N)$ necessary
to depin the hexagonal and the circular incommensurate islands made up of N atoms.
In alternative to that, to represent a commensurate case we also simulate the depinning of a model
now representing $\sqrt{3}$$\times$$\sqrt{3}$ islands, obtained with potentials of similar amplitudes
but no longer incommensurate and no longer at very low temperature (where kinetics would be too slow) but
at a reasonably high temperature near 50\,K (details in Methods).

Our main result is that the static friction of incommensurate islands is very small but nonzero, and interestingly area-dependent. 
For each given shape 
the static friction
obeys with striking precision a power law scaling $F_s \sim F_0 (A/\Sigma)^{\gamma_s}$
where $\Sigma = a^2 \sqrt{3}/2$ is the area per adsorbed atom.
The incommensurate exponent $\gamma_s$ varies between 0.25  and 0.37 depending on the choice of edge morphology, here polygonal
or circular, respectively. While of course real islands will generally look like neither, the two cases provide
reasonably  extreme instances, the circular probably more realistic than the polygonal, in virtue of its greater
variety of edge atom configurations.
We further show that 
in both shapes the island static friction barrier resides in a small set of points on the edge,
as a consequence of which the static friction exponent $\gamma_s$ is definitely less than 1/2, the value expected
if the density of the pinning points on the edge was uniform.

The commensurate island static friction obtained in simulation is by comparison five to six  orders of magnitude larger
than the incommensurate static friction, which demonstrates and emphasizes the huge gap between a strong and a weak contact.
With increasing island size, we find that the commensurate friction force $F_s$  grows proportionally to the area,
$\gamma_s = 1$, as expected when the bulk of the island participates in the static friction barrier, resulting in a
static friction stress $F_s /A$  independent of area.

The size-independent static friction shear stress $F_s/A$ for our model commensurate islands of small size is about
$10^7 N/m^2 $.  We can compare that with the nominal theoretical interface stress of a rare gas on metal (G/M) interface
$\mu_{G/M} = \frac{2 \mu_{G} \mu_{M}}{\mu_{G} + \mu_{M}}  \simeq1.16* 10^{9} N/m^2 $ (where $\mu \equiv  C_{44}$
are the respective bulk shear elastic constants, here of Kr and Pb ), to conclude that the commensurate island static friction stress obtained
is about $ 10^{-2} \mu_{G/M}$. Interestingly, that falls inside the range $\mu_{G/M}/30 \div \mu_{G/M}/1300$ expected
from dislocation theory of macroscopic, commensurate and inhomogeneous strong contacts. \cite{hurtadoI99, hurtadoII99, gao10}.

Returning to our main system of interest, the simulated incommensurate Kr/Pb islands, the shear stress $F_s/A$
is many orders of magnitude smaller than the commensurate, and unlike that case is clearly size-dependent. A physically relevant
island area where we wish to evaluate the static friction force $F_s(A)$ is the critical area at which it equals the peak inertial
force in a QCM experiment, $ F_{QCM}(A_{crit}) =\rho A_{crit} \Delta (2 \pi f )^2$, where $\rho$ is the adsorbate 2D mass density,
$\Delta$ the QCM oscillation amplitude, $f$ the frequency. Only in islands whose area exceeds $A_{crit}$ the
inertial force exceeds static friction, and inertial depinning 
followed by dynamic frictional sliding
can take place. Experimental QCM orders of magnitude being
$\Delta \sim$ 100\,\AA, $f \sim$ 10 MHz yield an inertial force which grows linearly with A and crosses,
as shown in Fig.~\ref{depforce},  the static friction lines  $F_s/A_{crit} = F_{QCM} /A_{crit}$, signaling
depinning above a critical area
\begin{equation}
A_{crit}  /\Sigma \sim \left({\frac{F_0}{\Sigma \rho \Delta (2 \pi f )^2}}\right)^{\frac{1}{1-\gamma}}.
\end{equation}
The incommensurate island data of Fig.~\ref{depforce}, obtained for 
an assumed 
5\% Pb(111) substrate corrugation,
yields $N_{crit} = A_{crit}/\Sigma \simeq 3\cdot 10^{9}$, or $A_{crit} \simeq  4.3\cdot 10^{10}$\,\AA$^2$.

That amounts to island diameters of the order of several microns,  corresponding to an inertial static friction force
of order 16 nN, equivalent to a QCM depinning stress $\simeq 40 N/m^2$, more than seven orders of magnitude smaller than
$\mu_{Kr/Pb}$,  and more than five orders of magnitude smaller than the commensurate island static friction
stress $\simeq 10^7 N/m^2$. Qualitatively similar data 
are found for lower corrugations, where 
the static friction force  
and thus the critical depinning area $A_{crit}$ 
drops by a factor $\sim 5\div 15$  when the corrugation is decreased from a large 5\% (where phenomena are easier to study) 
to $2\div 1 \%$.

At larger incommensurate
lattice mismatch, and under the effect of temperature the critical island size for depinning and the static
friction shear stress will be correspondingly smaller, but we still expect large critical island diameters ranging from many
tens to hundreds of nanometers.

Incommensurate islands smaller than this size will remain pinned by their own edge, and only larger ones will overcome
the edge-originated static friction and slide inertially in QCM. 
These large estimated critical radii constitute a
strong result of this work: the intrinsic edge-related static friction is by no means a small or academic effect, as
one might initially have expected.
Strong as it is, the edge contribution can in fact add significantly to impurity and defect pinning\cite{muser01},
even after renormalization of $F_s$ caused by edge roughness and temperature.
The order of magnitude obtained for the critical depinning island size 
must be compared with
the general experimental
observation of a critical adsorbate coverage of 5-30\% below which static friction wins,
there is no low temperature inertial depinning, and the slip time remains zero even on the cleanest surface,
and even for small oscillation amplitudes, a regime where dilute defects play a smaller role.
If a real surface is crudely assumed to consist of terraces of hundreds nanometers in size, the $A_{crit}$
corresponds to a non negligible submonolayer coverage, in agreement with experimental observations,
such as e.g., in Ne on Pb(111)\cite{fois07, bruschi10}.
For a strongly commensurate island conversely the inertial force $ F_{QCM}(A)$ and the static friction force $F_s$ are both proportional
to $A$ and do not cross as A grows, indicating that  the island will not inertially depin and remain stuck by its own bulk static friction,  
at least until larger island sizes and temperatures outside of this study.

\section{\bf {Edge pinning and soliton flow}}

We can now address the question, what is the physics of edge-induced pinning?
One key observation is that in order  for an island with lattice parameter $a$ to slide over the surface
with lattice spacing $b$, the misfit soliton lattice, of spacing $[\frac{a}{b}-1]^{-1}$,
must {\it flow} across the island in the sliding direction with a much larger speed $v [\frac{a}{b}-1]^{-1}$
than the overall island speed $v$ -- solitons must move very fast in order for the island to move even slowly (see Supplementary Movie 1).
The sublinear size scaling of $F_s$ with $\gamma_s < 1/2$ in Fig.~\ref{depforce}
occurs because solitons enter the island through specific points and not everywhere along the edge --
a corner in the perfect hexagonal shape for example as seen in the movie --
and then sweep past, setting the island's soliton pattern in a state of flow. 
The motion of solitons is necessary for the island to slide. By preventing the free motion of solitons,
which implies their free entry and  exit, the island edges cause static friction.
The local planar density of atoms for an edge-pinned island at pulling force just below the static friction threshold
shows an accumulation on the island front and a rarefaction on the back side (see Supplementary Figure 1),
also demonstrating the importance of non-rigid effects in larger regions surrounding the edge -- the pinning agent.
The edge adatoms lower their potential energy by settling
in local minima, thus breaking the ideal island-surface translational invariance.  
We note incidentally that the settling of edge atoms also implies a local vertical deformation, besides a horizontal
one, of the island boundary\cite{fasolino2013}. The overall relaxation 
gives rise to an edge related (``Peierls-Nabarro'') energy barrier for the motion of solitons
into and out of the island: the 2D edges cause the nominally incommensurate island to become
stuck and pinned against sliding. 

The observed depinning phenomenology with solitons getting in and out
of the island edge resembles that of the caterpillar-like motion of the finite one-dimensional (1D)
Frenkel-Kontorova (FK) model, where an edge barrier opposing kink motion gives rise to static friction\cite{braun_book}.
Many additional 2D factors, such as sliding-induced relative lattice orientation, island shape, 
vertical relaxations, edge rugosity, adsorbate
elasticity, etc., will enter in determining the exact value of the sublinear $\gamma_s$ exponent,
whose nontrivial theoretical description we do not attempt here.

\begin{figure}
\centering
\includegraphics[angle=0,width=8.5cm]{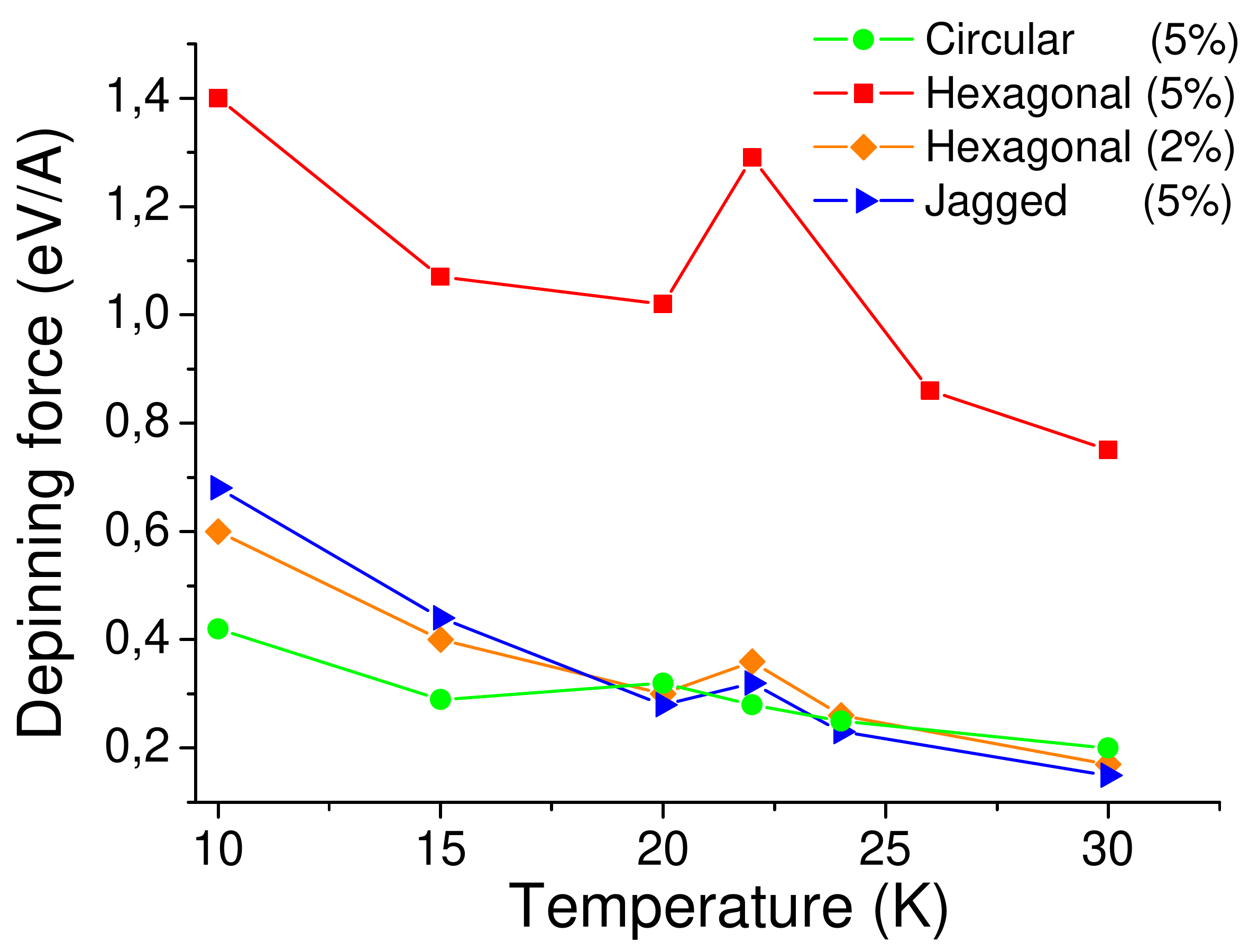}
\caption{\label{depinningtemp} {Static friction force $F_s$ as a function of temperature, for differently shaped islands (hexagonal, circular, and jagged, of about 3$\cdot$10$^5$ atoms)
and substrate energy corrugations (between 2 and 5 \% of the atom adsorption energy).
Superposed to the general decrease, the visible peak at 22 K reflects the low order adsorbate-surface
6/7 commensurability accidentally hit due to the island thermal expansion. This kind of accident,
here portrayed for our model of Kr/Pb(111), might more generally lead to re-entrant pinning in QCM.}}
\end{figure}
%

\section{\bf {Temperature effects}}

We come finally to discuss the effects of temperature on the edge-dependent static friction
of incommensurate islands. The thermal evolution of static friction provided by simulations is shown in Fig.~\ref{depinningtemp} for the
hexagonal, circular, and jagged islands (See Methods), and for different substrate energy corrugations.
Not surprisingly, thermal fluctuations facilitate the reduction of edge energy barriers, lowering static friction
and favoring island depinning under the action of $F_{ext}$.

A second temperature effect will be to enhance thermal roughening of the island's edge, via
adatom migration and 2D evaporation-condensation processes. The edge irregularity
slightly modifies the overall soliton map in the island, especially close to the edge atoms that are stuck,
a factor probably impacting the tribological response of the system, although most likely not more than
shown by the difference between hexagonal and circular islands.

\begin{figure}
\centering
\includegraphics[angle=0,width=8.5cm]{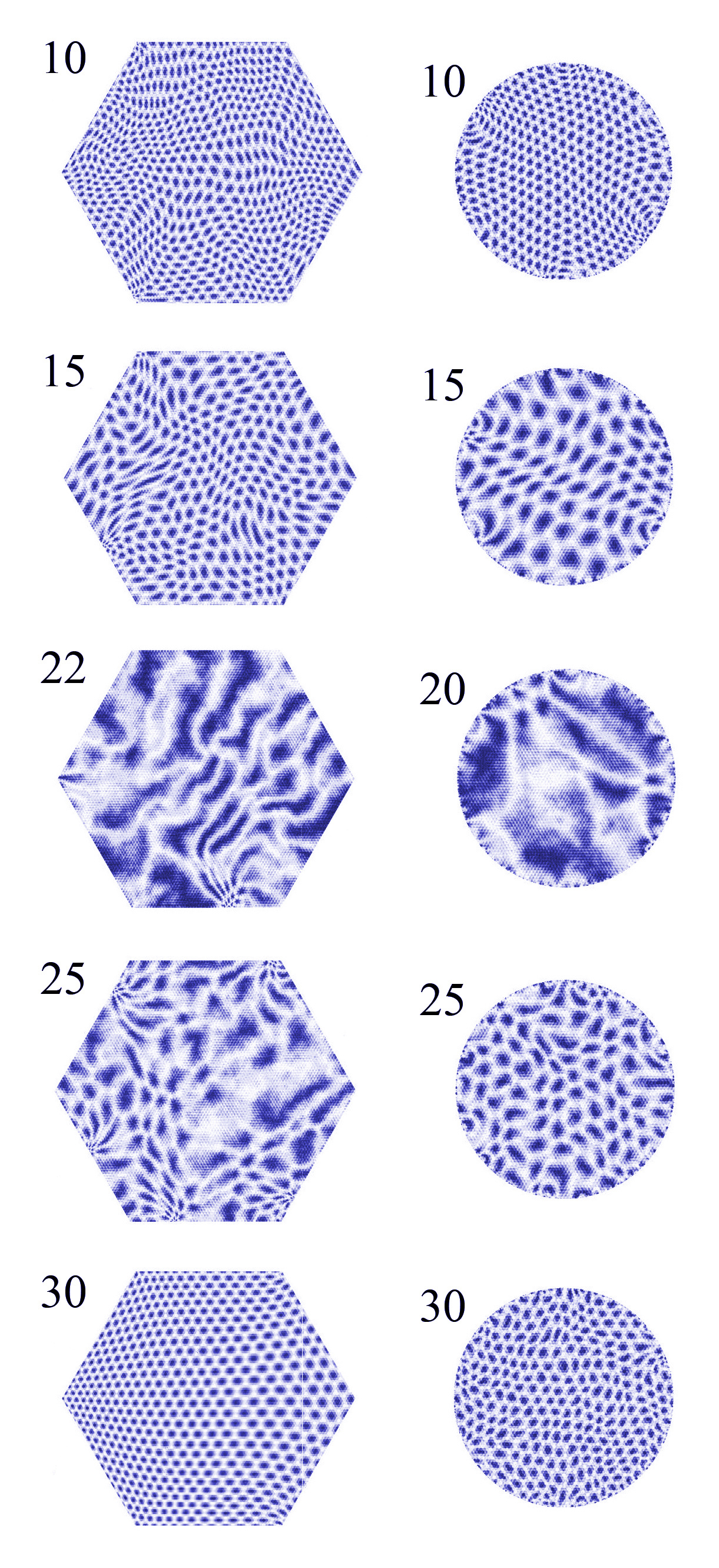}
\caption{\label{solitonskins}{Thermal evolution of the soliton network portrayed for hexagonal and circular shaped islands
using the contrast enhancement technique for the lateral density variations relative to a 6/7
commensurability (see Methods). The island thermally expands, going from relatively overdense at
low temperatures to underdense at high temperatures, thus exhibiting accidental near commensurability
around 20-22 K.}}
\end{figure}

A third temperature effect is the anharmonic thermal expansion of the island's 2D lattice
against the metal substrate, not yet expanding at these low temperatures.
This thermal drift of lattice parameter ratio between the two crystalline surfaces in contact
results in an alteration of the interface commensurability with consequences
on static friction. In the chosen example of Kr/Pb(111), the relative registry is expected to
drift from frank incommensurability at $10$\,K to a ``6/7'' higher order
commensurability near 20-25\,K and then again to incommensurability above that temperature.
In our simulations this accidental, weak high-order commensurability gives rise to a temporary rise
of static friction (Fig.~\ref{depinningtemp}).
Even if the effect is not strong, and is sensitively dependent upon unknowns such as the edge rugosity
and the substrate crystalline corrugation magnitude, in our realization we observe a small but visible
peak of the temperature-dependent static friction for all the simulated island shapes.
As highlighted in Fig.~\ref{solitonskins},
the corresponding smearing in the pattern of the soliton network, the
hallmark of an enhanced interface commensurability, for $T$\,=\,22\,K (hexagonal island)
and $T$\,=\,20\,K (circular island) is visible and pronounced.
This is suggestive of a more general possibility of expansion-induced reentrant static friction
peaking around a commensurate phase.

We note as an aside that since incommensurability relative to 6/7 switches from overdense below 22 \,K
to underdense above, the solitons are replaced in the process by antisolitons, entities which unlike solitons
flow in the opposite direction to the applied force. The antisolitons being essentially lines of vacancies in the 2D lattice,
their properties differ quantitatively from those of solitons, which are lines of interstitials.  Antisolitons are generally
of narrower width and less mobile, as seen for example in simulated sliding colloid monolayers\cite{Vanossi12PNAS}, another
system where some of the present results could in the future be verified -- although the nature and pinning role of boundaries
may be quantitatively different in that case. 
Although the resulting quantitative asymmetry between underdense islands (with stronger static friction) and overdense ones (with
weaker static friction) should not directly affect the critical exponents, it might do so indirectly by affecting differently 
the island edge shape and rugosity.
The above was for incommensurate islands. Very large at low temperatures the static friction of a commensurate island
 (see Methods) is crucially temperature dependent. It requires the thermal nucleation of a forward displaced domain.
The activation time $t_a$ for nucleation is expected to behave as in the edge-free bulk case\cite{zapperi, dellago13}
\begin{equation}
  t_a \sim C \exp(E_b/k_BT)
\end{equation}
where $E_b$ is the effective activation energy barrier, lowered by the applied external force, and dependent on the nucleation site.
As shown in Supplementary Movie 2, the island slides upon the temperature-related  appearance of the nucleus at the
island edge, followed by a subsequent force-driven expansion. Here once again the edge plays an important role,
even if different, in commensurate island static friction.  Nucleation at the edge implies that (owing to greater
atom mobility) there is a lower edge value E$_{b}^e$  of the barrier relative to the bulk barrier $E_b$ which
controls homogeneous nucleation\cite{zapperi}. Since $t_a$ establishes the threshold of depinning (see Methods),
one obtains for commensurate islands of increasing size a constant $t_a$, i.e. a constant $E_b$, only if the applied
force grows linearly with $N$, and therefore is $F_s \propto A$, or $\gamma_s$\,=\,1, exactly like in homogeneous nucleation.
We note incidentally that the same bulk-like static fricton scaling exponent $\gamma_s$\,=\,1  should also occur
for ``soft'' incommensurate islands, whose free sliding is dynamically unaccessible\cite{peyrard83}.
By contrast, for hard incommensurate islands the external force is resisted only by the edge pinning barrier, and
therefore $F_s$ will grow at most as $\sqrt{A}$ for a constant $E_{b}^e$. In real incommensurate islands
the inhomogeneous stress along the edge further reduces $E_b$
at specific points, causing the above-discussed sublinear size scaling of $F_s$ with $\gamma_s < 1/2$.
%

\section{\bf {Discussion and conclusions}}

We have conducted, in summary, a simulation study of atomistic sliding exhibiting static friction,
an important tribological parameter whose physics had been, despite specific case studies\cite{Sorensen96, hurtadoI99,
hurtadoII99, robbins1999, muser01, muser03, gao10, zapperi, dellago13}, still insufficiently addressed so far
for weak incommensurate contacts. With an eye to adsorbed rare gases typical of QCM experiments we simulated
the forced depinning of islands, and found that the island edges play an all-important role, blocking especially
the onset of otherwise superlubric sliding below a critical diameter. For rare gas islands inertially pushed on metal substrates,
the edge-originated static friction is relevant, hindering the low temperature sliding of islands whose interior would
otherwise be superlubric. Static friction is predicted for these islands to obey a sublinear
scaling growth with island area, with an exponent roughly between 0.25 and 0.37, depending on the island shape.
Previous work by Sorensen et al.\cite{Sorensen96}\ showed some effects due to edges of an AFM-like sliding of a Cu tip on a Cu surface.
More recently, the {\it dynamic} sliding friction, in principle a different quantity from static friction, was studied in nanomanipulation
experiments by Schirmeisen's group and found to obey a sublinear scaling laws\cite{dietzel13}.

That raises the conceptually interesting question of what should be the mutual relationship evolution of static and dynamic
friction
when size grows. This question is presently open, since it is hard to extract reliable dynamic friction size scaling
from simulation, whereas conversely the experiments cannot directly address the static friction and its scaling laws.
The lack of distinction found in some literature between static and dynamic friction remains in need of justification.
The dynamic friction force $\mathcal{F}$ of a hard incommensurate island will consist of $\mathcal{F}_b + \mathcal{F}_e$,
namely a bulk term plus an edge term. The edge term $\mathcal{F}_e$ should size-scale sublinearly like static friction
with an exponent smaller than 1/2, and will in case of stick-slip also imply a sublinear dependence upon center-of-mass
sliding speed. The bulk term $\mathcal{F}_b$ (vanishing in the static case) should conversely reflect superlubricity, with a linear, viscous speed
dependence, and a bulk-like size scaling exponent $\gamma$\,=\,1. For large speed and large island size the bulk dynamic
friction will eventually prevail, with a size dependence very different from that of static friction.
At much lower speed and at moderate size the bulk contribution should become less important, until eventually both static
and dynamic friction should become edge-dominated, with a similar sublinear size scaling.

It will be very interesting to watch a future expansion of experimental data to verify if static friction
will or not follow scaling laws we find here, with possible modifications when islands are replaced
by metal clusters with considerably more rigid edges and stronger contacts, or in the case of 2D colloids, where
edges may be completely different. The possible future use of QCM substrates with
controlled size terraces, obtained e.g.\ by vicinal surfaces, should enable a verification of the static friction
phenomena described here. The dependence of the static friction upon temperature is another result that could be
experimentally pursued, while an estimate of the substrate corrugation magnitude, a generally unknown parameter,
could be obtained by comparing data and simulations. More generally, the role of the edges of incommensurate sliding islands and clusters,
not sufficiently emphasized until recently, must be acknowledged as an important source of static friction.

\section{\bf{Computational Details}}

\subsection{Model interactions}

In our model the Pb(111) hard substrate is treated as a fixed and rigid triangular lattice frame, exerting on
the mobile Kr adatoms an average attractive potential $V$\,$\sim$\,-150 meV, and a weak
corrugation $\Delta V/V$ roughly in the range 2-5\% between the on-top
Pb site (the energy is minimum for a Kr adatom), and the hollow Pb site (energy maximum).
Each Kr adatom is thus submitted to the overall potential $V = V_{\mathrm{Kr-Kr}} + V_{\mathrm{Kr-Pb}}$.
The Kr-Kr interaction is modeled as regular Lennard-Jones (LJ) potential,
with $\epsilon$\,=\,0.014\,eV and $\sigma$\,=\,3.65\,\AA. Tiny corrections due to three-body forces
as well as substrate-induced modifications of this two-body force are ignored.
The Kr-Pb interaction is modeled by the following Morse-modified potential:
\begin{equation}
V_{\rm{Kr-Pb}} = \alpha(x,y)\left(\rm{e}^{-2 \beta(z-z_0)} - 2\rm{e}^{\beta(z-z_0)}\right).
\end{equation}
The (111) structure of the substrate is accounted for by replacing the function
$\alpha(x,y)$, that must exhibit the same periodicity of the underlying triangular lattice
(we neglect here the small differences between fcc and hcp stacking sites).
To represent the substrate modulation, we make use of the function
\begin{equation}
  \label{ModulatingFunction}
  M(x,y) = \frac{2}{3}
           - \frac{4}{9} \cos\left(\frac{2\pi x}{b}\right) \cos\left(\frac{2\pi y}{\sqrt{3} b}\right)
           - \frac{2}{9} \cos\left(\frac{4\pi y}{\sqrt{3} b}\right).
\end{equation}
The constant $b$ is the nearest neighbor distance of surface atoms.
The modulating function $M(x,y)$ has been normalized to span the interval from
0 (top sites) to 1 (hollow sites). Back now to the Morse potential,
the energy parameter is given by $\alpha(x,y) = \alpha^{top} + M(x,y) (\alpha^{hollow}-\alpha^{top})$
(see Supplementary for parameter details).
\\The LJ parameters leads to a nearest neighbor Kr-Kr distance which at 10\,K
(the lowest temperature of validity of our classical simulations, roughly equal to the
temperature of quantum freezing in Kr) is close to the 3D experimental value
of 4.01\,\AA, in turn 13\% higher than the Pb-Pb triangular substrate nearest
neighbor distance of 3.50\,\AA. Thus, the island and substrate 2D lattices are incommensurate
with a ratio of 0.8728 at $T$\,=\,10\,K. Upon moderate heating, the island readily expands
but the substrate does not. The closest strong commensuration of 6/7\,=\,0.8571 of Kr/Pb(111)
is reached and surpassed near $T \approx 20-22$\,K.
\\To model the commensurate islands we reparametrized the substrate potential
in order to match adhesion and corrugation energies of 190\,meV and 1.9\,meV, respectively,
and a lattice constant of 3.61/$\sqrt{2}$\,\AA, borrowing them from the Xe/Cu(111) case\cite{zapperi}.
Starting from a Xe-Xe Lennard-Jones energy $\epsilon=20$\,meV, we used an ad-hoc $\sigma=3.90$\,\AA~
in order to fictitiously match, at T\,=\,50\,K, the $\sqrt{3}$$\times$$\sqrt{3}$ Cu(111) spacing.

\subsection {Simulation procedure}
To simulate QCM depinning, we applied to all atoms of the annealed adsorbate island (see below)
a constant planar force. (This is adequate because the QCM oscillation period of $\sim$\,$10^{-7}$\,$s$ is much longer than our simulation times).
Static friction was measured in incommensurate systems using the following protocol:
(i) an initial 1.2 ns annealing run at $T$\,=\,25\,K, with no force applied ($F_{ext}=0$);
(ii) a 1.2 ns run at $T$\,=\,0\,K, with typical Berendsen thermostat constant $\tau$\,=\,100 ps and still $F$\,=\,0;
(iii) the thermostat is removed, and a series of 1.2 ns runs at 0\,K, with $F$ applied instantaneously after the annealing procedure.

Simulations of commensurate systems were performed at $T$\,=\,50\,K. Similarly to above, we increased the external force from zero to $F_s$,
at increments of 0.01\,meV/atom, letting the simulation evolve for 100\,ps after each force step.
In all cases, a displacement of the island center-of-mass by 2 lattice spacing within the same force step is taken as the signal of depinning.
The static friction so obtained should depend slightly on the waiting time, but we verified
that the decrease due to thermal barrier hopping was negligible with a waiting time  longer than 1.2 ns.

At sufficiently low temperature and small size, the adsorbate atoms form spontaneously
a triangular lattice weakly distorted by the underlying substrate.
With the realistic choice $\sigma$\,=\,3.65\,\AA~for the LJ potential of
krypton, the ideal 2D lattice has a spacing $a$\,=\,4.0568\,\AA.
Exact $6/7$ commensurability would require  the slightly larger value
4.0842\,\AA. Due to this mismatch, the adsorbate island shows commensurate domains joined by bands
of higher atomic density: the solitons, whose exact pattern depends
on parameters, such as the temperature and the island size.
To visualize the soliton superstructures, we developed a method exploiting
information from all the adsorbed atoms. If the krypton atoms would occupy
the position of a perfect triangular lattice with commensurability ratio $6/7$,
a uniform dilation by a factor 6 would map each adatom onto a position on top of  a substrate atom.
Thus, given an atom with coordinates $(x,y,z)$, its dilated position $(X,Y,Z)$\,=\,$(6x,6y,6z)$
must  render the modulating function $M$ (Eq. \ref{ModulatingFunction}) equal to zero, in the commensurate domains.
The solitons, characterized by an enhanced mobility due to the mismatch with the substrate,
are instead regions with $M(6x,6y)$\,$\gtrsim$\,0.
Plotting each atom in bluescale, ranging from blue to white
according to increasing values of $M(6x,6y)$, the commensurate domains appear dark blue and the
solitons as white bands.\\

\section {\bf{Acknowledgements}}
We especially acknowledge early collaboration with
U. Tartaglino and F. Ercolessi, and helpful discussions with N. Manini.
The work was partly funded by
the Swiss National Science Foundation through a SINERGIA contract CRSII2\_136287, by PRIN/COFIN  Contract 2010LLKJBX 004,
by COST Action MP1303, and mainly by the ERC Advanced Grant No. 320796-MODPHYSFRICT.
The CINECA supercomputing center is also gratefully acknowledged. \\

\footnotesize{

\begin{thebibliography}{999}

\bibitem{bhushan2002} B. Bhushan, J.N. Israelachvili and U. Landman,
{\it Nature}, 2002, {\bf 374}, 607.

\bibitem{urbakh2004} M. Urbakh, J. Klafter, D. Gourdon and J.N. Israelachvili,
{\it Nature}, 2004, {\bf 430}, 525.

\bibitem{urbakhmeyer} M. Urbakh and E. Meyer,
{\it Nature Mater.}, 2010, {\bf 9}, 8.

\bibitem{vanossi2013} A. Vanossi, N. Manini, M. Urbakh, S. Zapperi and E. Tosatti,
{\it Rev. Mod. Phys.}, 2013, {\bf 85}, 529.

\bibitem{frenken2006} J.W.M. Frenken,
{\it Nature Nanotech.}, 2006, {\bf 1}, 20.

\bibitem{schirmeisen08} D. Dietzel, C. Ritter, T. M\"onninghoff, H. Fuchs, A. Schirmeisen and U.D. Schwarz,
{\it Phys. Rev. Lett.}, 2008, {\bf 101}, 125505.

\bibitem{dietzel13} D. Dietzel, M. Feldmann, U.D. Schwarz, H. Fuchs and A. Schirmeisen,
{\it Phys. Rev. Lett.}, 2013, {\bf 111}, 235502.

\bibitem{kawai2014} S. Kawai et al.,
{\it Proc. Natl. Acad. Sci. USA}, 2014, {\bf 111}, 3968.

\bibitem{allison2009} H. Hedgeland, P. Fouquet, A.P. Jardine, G. Alexandrowicz, W. Allison and J. Ellis,
{\it Nature Phys.} 2009, {\bf 5}, 561.

\bibitem{mistura0}  A.Carlin,  L. Bruschi,  M. Ferrari, G. Mistura,
{\it Phys. Rev.},  2003, {\bf 68}, 045420.

\bibitem{krim_review} J. Krim,
{\it Nano Today}, 2007, {\bf 2}, 38.

\bibitem{krim_advphys} J. Krim,
{\it Adv. Phys.}, 2012, {\bf 61}, 155.

\bibitem{junno} T. Junno, K. Deppert, L. Montelius and L. Samuelson,
{\it Appl. Phys. Lett.}, 1995, {\bf 66}, 3627.

\bibitem{sitti} M. Sitti and H. Hashimoto,
{\it IEEE-ASME Trans. Mechatron.}, 2000, {\bf 5}, 199.

\bibitem{hurtadoI99} J. A. Hurtado and K.-S. Kim, {\it Proc. R. Soc. Lond. A}, 1999, {\bf 455}, 3363.

\bibitem{hurtadoII99} J. A. Hurtado and K.-S. Kim, {\it Proc. R. Soc. Lond. A}, 1999, {\bf 455}, 3385.

\bibitem{gao10} Y. Gao, {\it J. Mech. Phys. Solids}, 2010, {\bf 58}, 2023.

\bibitem{robbins1999} G. He, M.H. Muser and M.O. Robbins,
{\it Science}, 1999, {\bf 284}, 1650.

\bibitem{muser01} M.H. M\"user, L. Wenning and M. O. Robbins,
{\it Phys. Rev. Lett.}, 2001, {\bf 86}, 1295.

\bibitem{muser03} M.H. M\"user, M. Urbakh and M.O. Robbins,
{\it Adv. Chem. Phys.}, 2003, {\bf 126}, 187.

\bibitem{persson_book} B.N.J. Persson,
{\it Sliding Friction: Physical Principles and Applications}, Springer, 1998.

\bibitem{zapperi} M. Reguzzoni, M. Ferrario, S. Zapperi and C. Righi,
{\it Proc. Natl. Acad. Sci. USA}, 2010, {\bf 107}, 1311.

\bibitem{dellago13}
J. Hasnain, S.Jungblut and C. Dellago,
{\it Soft Matter}, 2013, {\bf 9}, 5867.

\bibitem{peyrard83} M. Peyrard and S. Aubry,
{\it J. Phys. Condens. Matter}, 1983, {\bf 16}, 1593.

\bibitem{hirano90} M. Hirano and K. Shinjo,
{\it Phys. Rev. B}, 1990, {\bf 41}, 11387.

\bibitem{shinjo93} K. Shinjo and M. Hirano,
{\it Surf. Sci.}, 1993, {\bf 283}, 473.

\bibitem{cieplak94} M. Cieplak, E.D. Smith and M.O. Robbins,
{\it Science}, 1994, {\bf 265}, 1209.

\bibitem{braun_book} O.M. Braun,
{\it The Frenkel-Kontorova Model: Concepts, Methods, and Applications}, Springer, 2004.

\bibitem{note1} The small static friction at the sliding interface between
two ideally hard disordered surfaces, e.g., metal on glass has also been discussed\cite{muser01,superlubricity_book}.

\bibitem{superlubricity_book} A. Erdemir and J.-M. Martin,
{\it Superlubricity}, Elsevier Science Amsterdam, 2007.

\bibitem{superlubric_china} R. Zhang, Z. Ning, Y. Zhang, Q. Zheng, Q. Chen, H. Xie, Q. Zhang, W. Qian and F. Wei,
{\it Nature Nanotech.}, 2013, {\bf 8}, 912.

\bibitem{superlubric_bocquet} A. Nigues, A. Siria, P. Vincent, P. Poncharal and L. Bocquet,
{\it Nature Mater.}, 2014, {\bf 13}, 688.

\bibitem{dienwiebel04} M. Dienwiebel, G.S. Verhoeven, N. Pradeep, J.W. Frenken, J.A. Heimberg and H.W. Zandbergen,
{\it Phys. Rev. Lett.}, 2004, {\bf 92}, 126101.

\bibitem{dienwiebel08} A. Filippov, M. Dienwiebel, J.W. Frenken, J.Klafter and M. Urbakh,
{\it Phys. Rev. Lett.}, 2008, {\bf 100}, 046102.

\bibitem{fasolino2013} M. M. van Wijk, M. Dienwiebel, J. W. M. Frenken and A. Fasolino,
{\it Phys. Rev. B}, 2013, {\bf 88}, 235423.

\bibitem{schirmeisen09} A. Schirmeisen and U. D. Schwarz,
{\it Chem. Phys. Chem.}, 2009, {\bf 10}, 2373.

\bibitem{Bohlein12} T. Bohlein, J. Mikhael and C. Bechinger,
{\it Nature Mater.}, 2012, {\bf 11}, 126.

\bibitem{Vanossi12PNAS} A. Vanossi, N. Manini and E. Tosatti,
{\it Proc. Natl. Acad. Sci. USA}, 2012, {\bf 109}, 16429.

\bibitem{Vanossi12} A. Vanossi and E. Tosatti,
{\it Nature Mater.}, 2012, {\bf 11}, 97.

\bibitem{zeppenfeld92} P. Zeppenfeld, U. Becher, K. Kern and G. Comsa,
{\it Phys. Rev. B}, 1992, {\bf 45}, 5179.

\bibitem{Bruch07}  L. W. Bruch, R. D. Diehl and J. A. Venables,
{\it Rev. Mod. Phys.}, 2007, {\bf 79}, 1381.

\bibitem{mistura1} L. Bruschi, A. Carlin and G. Mistura,
{\it Phys. Rev. Lett.}, 2002, {\bf 88}, 046105.

\bibitem{mistura2} L. Bruschi, G. Fois, A. Pontarollo, G. Mistura, B. Torre, F. Buatier de Mongeot, C. Boragno, R. Buzio and U. Valbusa,
{\it Phys. Rev. Lett.}, 2006, {\bf 96}, 216101.

\bibitem{bruschi10} L. Bruschi, M. Pierno, G. Fois, F. Ancilotto, G. Mistura, C. Boragno, F. Buatier de Mongeot and U. Valbusa,
{\it Phys. Rev. B.}, 2010, {\bf 81}, 115419.

\bibitem{Park99} Ji-Yong Park et al.,
{\it Phys. Rev. B}, 1999, {\bf 60}, 16934.

\bibitem{renner01} R.L. Renner, J.E. Rutledge and P. Taborek,
{\it Phys. Rev. B}, 2001, {\bf 63}, 233405.

\bibitem{fois07} G. Mistura, private communication.

\bibitem{kern87} K. Kern, P. Zeppenfeld, R. David and G. Comsa,
{\it Phys. Rev. Lett.}, 1987, {\bf 59}, 79.

\bibitem{QCMnote} Note that QCM oscillations, of typical frequency $f_0 \sim 10$\,MHz,
are essentially static on the simulation timescales.

\bibitem{Sorensen96}
M. R. Sorensen, K. W. Jacobsen and P. Stoltze,
{\it Phys. Rev. B}, 1996, {\bf 53}, 2101.

\end{thebibliography}

\bibliographystyle{rsc} 
}

\end{document}